\documentclass[12pt]{article}

\begin{document}

\title{Virial coefficients from 2+1 dimensional QED effective 
actions at finite temperature and density}
\author{P. F. Borges$^1$\thanks{%
e-mail: pborges@cefet-rj.br}, H. Boschi-Filho$^2$\thanks{ 
e-mail: boschi@if.ufrj.br}\, and Marcelo Hott$^3$\thanks{ 
e-mail: hott@feg.unesp.br} \\
{\small \textit{(1) Centro Federal de Educa\c c\~ao Tecnol\'ogica 
Celso Suckow da Fonseca }}\\
{\small \textit{Coordena\c c\~ao de F\'{\i}sica, Av. Maracan\~a, 229,
Maracan\~a}}\\
{\small \textit{20271-110 Rio de Janeiro, BRAZIL}}\\
{\small \textit{(2) Instituto de F\'\i sica, 
Universidade Federal do Rio de Janeiro }}\\
{\small \textit{Cidade Universit\'aria, Ilha do Fund\~ao, 
Caixa Postal 68528 }}\\
{\small \textit{21941-972 Rio de Janeiro, BRAZIL}}\\
{\small \textit{(3) Departamento de F\'\i sica e Qu\'\i mica, 
Universidade Estadual Paulista}}\\
{\small \textit{Campus de Guaratinguet\'a, Caixa Postal 205}}\\
{\small \textit{12500-000 Guaratinguet\'a, S\~ao Paulo, BRAZIL}}}
\date{}
\maketitle

\begin{abstract}
From spinor and scalar 2+1 dimensional QED effective actions at 
finite temperature and density in a constant magnetic 
field background, we calculate the corresponding virial coefficients 
for particles in the lowest Landau level. 
These coefficients depend on a parameter $\theta $ 
related to the time-component of the gauge field, 
which plays an essential role for large gauge invariance. 
The variation of the parameter $\theta$ might lead to an interpolation 
between fermionic and bosonic virial coefficients, although these
coefficients are singular for $\theta=\pi/2$.
\end{abstract}

\baselineskip = 20pt

\bigskip 
\noindent PACS number(s): 11.10.Wx, 05.30.Pr, 12.20.-m. 
\par

\newpage

\section{Introduction}

Gauge invariance in 2+1 spacetime dimensions allows the existence 
of a mass term for gauge fields, known as the Chern-Simons (CS) 
term \cite{CS}. This term can be generated dynamically by radiative 
corrections \cite{redlich84} but can not preserve parity and gauge 
invariance simultaneously. At finite temperature, gauge invariance 
takes on a particular form being large (not infinitesimal) with 
respect to the compactified Euclidean time, leaving a free parameter 
in the theory, usually a constant value of the time-component
of the gauge field. Recently, it has been shown that a resummation 
of the one-loop graphs is essential to preserve the large gauge 
invariance of the finite temperature 2+1 dimensional quantum 
electrodynamics (QED$_3$) effective action \cite{dunne97,schapo97}.

In this letter we compute explicitly the virial coefficients from 
spinor and scalar QED$_3$ effective actions in a constant magnetic 
field background for particles in the lowest Landau level, 
maintaining large gauge invariance.

The virial coefficients obtained here are similar to those found 
previously \cite{bbf123} imposing generalized boundary conditions 
to scalar and spinor fields in 2+1 dimensions in the absence of 
magnetic fields. The possibility of relating these boundary 
conditions to fractional statistics was envisaged in a quantum 
mechanical one-dimensional model \cite{bfd}. It has also been
shown in different contexts at finite temperature that if one 
considers an imaginary part of the chemical potential, fermions 
can transmutate into bosons \cite{barbosa}.

In fact, the coefficients found here depend on the time-component 
of the gauge field which can be seen as an imaginary part of the 
chemical potential and whose variation might lead to an interpolation 
between fermionic and bosonic coefficients. However, such 
coefficients differ from those coming from non-relativistic 
anyons (quasi-particles with fractional statistics) 
\cite{leinaas77,arovas85,JC90,kim,VO94} and are singular for a 
certain combination of the time-component of the gauge field and 
the temperature.

The physical picture that would be responsible for generating
the interpolation of the bosonic and fermionic virial coefficients 
for both scalar and spinor QED$_3$ at finite temperature and density 
is the large gauge invariance together with the cylindrical structure 
of the space-time, $\mathcal{R}^2 \times \mathcal{S}^1$. 
This non-trivial topology inhibits the removal of the time component 
of the gauge field leaving a free parameter in the theory.

%%%%%%%%%%%%%%%%%%%%%%%%%%%%%%%%%%%%%%%%%%%%%%%%%%%%%%%%%%%%%%%                                                                              

\section{Effective actions at finite temperature and density}

%%%%%%%%%%%%%%%%%%%%%%%%%%%%%%%%%%%%%%%%%%%%%%%%%%%%%%%%%%%%%%%

In QED the one-loop effective action for the gauge field 
$A_\nu=A_\nu(x)$ is obtained by integrating out the fermion field. 
In 2+1 dimensions one has 
\begin{eqnarray*}
\exp\left\{i\mathcal{S}_{eff}[A_\nu]\right\} 
=\int\mathcal{D}\psi\mathcal{D}
\bar{\psi} \exp\left\{i\int d^{3}x \bar{\psi} 
[\gamma^\nu(i \partial_\nu+e{A}
_\nu)-m] \psi- \frac{1}{4}F^{\rho \nu}F_{\rho \nu}\right\}.
\end{eqnarray*}

\noindent The effective action can be calculated exactly for some
configurations of the gauge field in which cases the field is 
understood as a classical background.

The thermodynamics of a system in thermal equilibrium at a finite
temperature $T=\beta^{-1}$ is described by its partition function 
$\mathcal{Z}$ which can be obtained by a Wick rotation of the 
effective action from Minkowski to Euclidean space with the 
imaginary time $\tau=-ix^0$ compactified into the interval 
$[0,\beta]$. Further, to describe the system at finite density 
we introduce a chemical potential $\mu$ which is identified 
with the imaginary part of the time component of gauge field 
$A_\nu$. The free energy 
$$\Omega(\beta,\mu)\; =\; - \; \frac{1}{\beta}\;\ln \mathcal{Z}$$
within this prescription \cite{kapusta} is given in terms of 
the real part of the effective action as 
$$\Omega(\beta,\mu)\; =\; - \; \frac{1}{\beta}\; 
\Re\{\mathcal{S}_{eff}(\beta,\mu)\}.$$ 
It can be expanded in terms of the fugacity $z=\exp(e\beta\mu)$ as 
\begin{eqnarray}
\Omega(\beta,\mu) \; =\; -\; {\frac{S}{\beta}}\; \sum_{n}\, b_n z^n\,,
\label{Omegabn}
\end{eqnarray}

\noindent where $S$ is the area and $b_{n}\equiv b_{n}(\beta )$ 
are the cluster coefficients. 
The pressure can be written as a power expansion of
the density as 
$$P\;\beta\; =\; \sum_{n=1}^{\infty }\; a_{n}\;\rho ^{n},$$ 
where $a_{n}$ are the virial coefficients which can be obtained 
through the standard relations \cite{Dash}: 
\[
b_{1}=a_{1}b_{1}; 
\]
\[
b_{2}=2a_{1}b_{2}+a_{2}{b_{1}}^{2}; 
\]
\[
b_{3}=3a_{1}b_{3}+4a_{2}b_{1}b_{2}+a_{3}{b_{1}}^{3}; 
\]
\begin{equation}
b_{4}=4a_{1}b_{4}+4a_{2}({b_{2}}^{2}+6b_{1}b_{3})+6a_{3}{b_{1}}
^{2}b_{2}+a_{4}{b_{1}}^{4};  \label{ca}
\end{equation}
\[
\vdots 
\]

%%%%%%%%%%%%%%%%%%%%%%%%%%%%%%%%%%%%%%%%%%%%%%%%%%%                                                                      

\section{The virial coefficients from fermions}

%%%%%%%%%%%%%%%%%%%%%%%%%%%%%%%%%%%%%%%%%%%%%%%%%%%

Here we consider the QED$_3$ effective action for a configuration 
where the time-component of the gauge field depends only on the 
Euclidean time $A_{3}= A_{3}(\tau)$, and choose 
$A_{j}= \frac{1}{2} F_{jk} x^{k}$ with $j=1,2$
corresponding to a constant magnetic field background ($F_{12}=B$). 
In this case it is possible to obtain an exact result for the 
fermion propagator, the effective action 
\cite{redlich84}, and generate the CS term dynamically preserving 
large gauge invariance \cite{schapo97}. Further, we work with a 
($2\times 2$) irreducible representation of Dirac matrices in
2+1 dimensions implying a CS term that explicitly breaks parity 
invariance.
Besides, there is a parity invariant contribution that comes from 
the Maxwell term which in this case corresponds to a constant 
magnetic field $B$.

The effective action at finite temperature and density that comes 
from the parity violating Lagrangian density can be found to be 
\cite{felipe98,HottMetikas} 
\begin{eqnarray}
\mathcal{S}_{eff}^{PV} (\beta,\mu) 
= \frac{e B S}{4 \pi } \frac {m}{|m|} 
\Big\{ i e \beta\,\Xi + G_+(|m|+ie \,\Xi) - G_+(|m|-ie \,\Xi) \Big\},
\label{SeffPV}
\end{eqnarray}
\noindent where we defined $$G_+(x)=\ln(1+e^{-\beta x})$$ 
and included the chemical potential contribution through 
$\,\Xi={\tilde A}_{3}+i\mu$ where $\tilde{A}_{3}$ 
is a \textsl{constant} related to $A_{3}(\tau )$ by the large
gauge transformations \cite{schapo97} 
\begin{eqnarray}
\tilde{A}_{3}=\frac{2\pi k}{e\beta}+\frac{1}{\beta }
\int_{0}^{\beta }d\tau
A_{3}(\tau )\,,  \label{A3tilde}
\end{eqnarray}

\noindent with $k=0,\pm 1,\pm 2,...$ being the winding number. 
Since the finite temperature and density QED effective actions 
that we discuss here depend on ${A}_3(\tau)$ through 
$\tilde{A}_3$ they are gauge invariant under
large gauge transformations.

The effective action at finite temperature and density that comes 
from parity invariant Lagrangian can be written as \cite{HottMetikas} 
\begin{eqnarray}
\mathcal{S}_{eff}^{PI} (\beta,\mu) = \frac{e |B| S}{4 \pi } 
\sum_{\ell=0}^{\infty} \sum_{s=1}^{2} \Big\{ \beta E_{\ell ,s} 
+ G_+(E_{\ell ,s} + ie\,\Xi) + G_+(E_{\ell ,s}-ie \,\Xi) \Big\}
\;,\label{SeffPI}
\end{eqnarray}
\noindent where $E_{\ell ,s}=\sqrt{m^{2}+2e|B|(\ell +s-1)}$ is the 
energy of the Landau levels.

%%%%%%%%%%%%%%%%%%%%%%%%%%%%%%%%%%%%%%%%%%%%%%
%\subsection{The Parity variant contribution}
%%%%%%%%%%%%%%%%%%%%%%%%%%%%%%%%%%%%%%%%%%%%%%

Now, we can calculate the cluster and virial coefficients from the 
above spinor effective actions. Expanding $G_+(x)$ as a power series 
of the fugacity $z$, we have that the corresponding contribution 
of the parity violating (PV) part Lagrangian to the free energy 
is given by: 
\begin{eqnarray}  \label{OmegaPV}
\Omega^{PV}(\beta,\mu) = {\frac{eBS }{2\pi\beta}}{\frac{m }{|m|}} 
\Big\{e\beta\mu + \sum_{n=1}^{\infty}{\frac{(-1)^n}{n}}e^{-n\beta|m|} 
\cos(en\beta \tilde{A}_3) \sinh(en\beta\mu)\Big\}\;.
\end{eqnarray}

\noindent The corresponding cluster coefficients can be obtained 
by comparing this expression with the expansion (\ref{Omegabn}). 
Note that in a non-relativistic system one usually finds only 
positive (or negative) powers of the fugacity $z$. 
However, in a relativistic system due to the presence of particles 
and antiparticles one finds both positive and negative powers of $z$, 
as in the above equation. Also the chemical potentials for particles
and antiparticles differ only by its sign, as a consequence of charge
conservation \cite{haberweldon}. The cluster coefficients for the PV
Lagrangian are then given by: 
\begin{equation}
b_{\pm n}^{PV} 
= \mp {\frac{(-1)^n}{n}}{\frac{eBm }{4\pi|m|}}e^{-n\beta|m|}
\cos(en\beta \tilde{A}_3)\;,  \label{bnPV}
\end{equation}
\noindent where $n=1,2,...$, the signs $\pm$ refers to particles and
antiparticles respectively, with opposite cluster coefficients while 
$b_0=-{\ e^{2}B\beta\mu m/4\pi |m|}$ does not contribute to the 
thermodynamics of the system once it depends linearly on $\beta$.

%%%%%%%%%%%%%%%%%%%%%%%%%%%%%%%%%%%%%%%%%%%%%%%%%
%\subsection{The Parity invariant  contribution}
%%%%%%%%%%%%%%%%%%%%%%%%%%%%%%%%%%%%%%%%%%%%%%%%%

The action that comes from the parity invariant (PI) Lagrangian 
includes the summation over all Landau levels. 
It can be split into the lowest Landau level (LLL) contribution 
$(E_{0,1}=|m|)$ and the excited states ones as 
\begin{eqnarray}
\Omega^{PI} (\beta,\mu) &=& - {\frac{e|B|S }{4\pi\beta}} 
\Re \bigg\{ \beta
|m|+ G_+(|m|+ie \,\Xi) + G_+(|m|-ie \,\Xi)  \nonumber \\
&&\;+ 2 \sum_{k=1}^\infty \beta E_{k} + G_+(E_{k}+ie \,\Xi) 
+ G_+(E_{k}-ie\,\Xi) \bigg\}\;,
\end{eqnarray}

\noindent where $E_{k}=\sqrt{m^2+2e|B|k}$. Then, the cluster 
coefficients in this case can be written as 
$$b_n^{PI}=b_n^{PI}|_{LLL} + b_n^{PI}|_{excited}$$
and the LLL case follows similarly to the PV case so that 
\begin{eqnarray}
{b_{\pm n}^{PI}}\vert_{LLL} = -{\frac{(-1)^{n}e|B|}{4n\pi}}
e^{-n\beta|m|}\cos(en\beta \tilde{A}_3)\; ,  \label{bnPILLL}
\end{eqnarray}

\noindent where $n=1,2,...$, and $\pm$ refers to particles and
antiparticles. In this case they have the same cluster coefficients 
and $b_0^{PI}|_{LLL}=-{e\beta |m B| / 4\pi}$ again does not affect 
the thermodynamics since it also depends linearly on $\beta$.

%%%%%%%%%%%%%%%%%%%%%%%%%%%%%%%%%%%%%%%%%%%%%%%%%%%%%%%%%%%%
%\subsection{The virial Coefficients in the Fermionic case}
%%%%%%%%%%%%%%%%%%%%%%%%%%%%%%%%%%%%%%%%%%%%%%%%%%%%%%%%%%%%

Now we can find the virial coefficients corresponding to the 
sum of PI and PV actions. Assuming that the magnetic field is 
strong and the temperature low enough to keep the particles 
in the LLL state we can take the contributions from the cluster 
coefficients eqs. (\ref{bnPV}), (\ref{bnPILLL}), so that 
(choosing $B>0$) we have 
\begin{eqnarray*}
b_{\pm n}^{LLL} = -{\frac{(-1)^{n}eB}{4n\pi}}(1\pm\frac {m}{|m|})
e^{-n\beta|m|}\cos(en\beta \tilde{A}_3)\;.
\end{eqnarray*}

\noindent 
Up to this point the distinction of particles and antiparticles 
($\pm $) is completely arbitrary. However, due to the factor 
$1\pm {m}/{|m|}$ one can see that only particles \textsl{or} 
antiparticles will contribute depending on the sign of $m$ 
that we choose. This result is closely related to the choice 
of the non-equivalent irreducible representation of Dirac
matrices that we pick up for the PV action. Have we chosen the 
other inequivalent irreducible representation we would have 
found the PV action with reversed signs so that the role of 
particles and antiparticles would be reversed. 
From now on we take $m>0$ and $n>0$, so that 
\begin{equation}
b_{n}^{LLL}
=-(-1)^{n}{\frac{\rho _{L}}{n}}e^{-n\beta m}\cos (n\theta )\;,
\label{bnLLL}
\end{equation}

\noindent where $\rho_L =2 \beta \omega_c / \lambda^2$ with 
$\omega_c=|eB|/2m$ being half of the cyclotron frequency and 
$\lambda=\sqrt{2\pi\beta/m}$ the thermal wavelength. 
Furthermore, we have defined $\theta\equiv e\beta \tilde{A}_3$, 
in analogy with the situation without the magnetic field 
background \cite{bbf123}. 
One can also define a non-relativistic chemical potential in
terms of the relativistic one up to the rest mass $m$ 
(in natural units $\hbar=c=1$). 
However, this redefinition of the chemical potential does not
affect the virial coefficients.

If we substitute the above cluster coefficients into relations 
(\ref{ca}), we are able to find all virial coefficients. 
Then, we obtain the virial coefficients 
\begin{eqnarray}  
\label{a2}
a_{2}^{LLL} &=&\frac{1}{2\rho _{L}}[1-\tan ^{2}\theta ]; \\
a_{3}^{LLL} &=&\frac{1}{3(\rho _{L})^{2}}[1+3\tan ^{4}\theta ]; \\
a_{4}^{LLL} &=&\frac{1}{4(\rho _{L})^{3}}[1-3\tan ^{4}{\theta }
+10\tan ^{6}{\ \theta }]\;; \\
a_{5}^{LLL} &=&\frac{1}{5(\rho _{L})^{4}}[1+20\tan ^{6}{\theta }
+35\tan ^{8}{\ \theta }]\;; \\
a_{6}^{LLL} &=&{\frac{1}{6(\rho _{L})^{5}}}[1-10\tan ^{6}\theta 
-105\tan^{8}\theta +126\tan ^{10}\theta ]\;.\label{a6}
\end{eqnarray}

\noindent Note that the above expressions for the virial 
coefficients diverge for $\theta ={\pi }/2$ 
and we have not found any apparent reason why it
should be so.

%%%%%%%%%%%%%%%%%%%%%%%%%%%%%%%%%%%%%%%%%%%%%%%%%

\section{The virial coefficients from bosons}

%%%%%%%%%%%%%%%%%%%%%%%%%%%%%%%%%%%%%%%%%%%%%%%%%

Now we turn to scalar QED$_3$. The effective action for the gauge 
field $A_\nu$ in this case can be calculated using the same choice 
we did (a constant magnetic field) for the fermionic case. 
Here, there is no parity violating Lagrangian once it does not 
involve Dirac matrices. Then the bosonic effective action is similar 
to the PI part of the fermionic case and we find 
\begin{eqnarray}  \label{Seffbos}
\mathcal{S}_{eff}^{B} (\beta,\mu) = \frac{eSB}{2\pi} 
\sum_{l=0}^{\infty} 
\Big\{ \beta E_{l}+G_-(E_{l}-ie \,\Xi) + G_-(E_{l}+ie \,\Xi)\Big\}\;,
\end{eqnarray}

\noindent where the energy of Landau levels is now given by 
$E_{l}=\sqrt{m^2+2eB(l+1/2)}$ and we defined 
$$G_-(x)=\ln(1-e^{-\beta x})\;.$$ 
As in the PI part of the fermionic 
case we can split the contribution of LLL state which
is given by $E_0=\sqrt{m^2+eB}$ from the excited ones, so that 
\begin{eqnarray}
\Omega^B (\beta,\mu) &=& -\, {\frac{eBS }{2\pi\beta}} 
\Re \Big\{ \beta E_0 + G_-(E_0-ie\,\Xi) + G_-(E_0+ie\,\Xi)  
\nonumber \\
&& +\, \sum_{l=1}^\infty \beta E_{l} + G_-(E_{l}-ie\,\Xi) +
G_-(E_{l}+ie\,\Xi) \Big\}\;.
\end{eqnarray}

\noindent Thus, the cluster coefficients can be written as 
$$b_{n}^{B}=b_{n}^{B}|_{LLL}+b_{n}^{B}|_{excited}\;.$$ 
Note that here, in opposition to the fermionic case, 
the LLL cluster coefficients depend on the magnetic field $B$ 
so we use the approximation 
$$E_{0}=\sqrt{m^{2}+eB}\approx m+{eB/2m}\;,$$ 
valid when $eB<<m^{2}$. 
Then, we find that the cluster coefficients corresponding to the 
LLL in this case are given by 
\begin{equation}
{b_{\pm n}^{B}}|_{LLL}={\frac{\rho _{L}}{n}}
e^{-n\beta (m+\omega _{c})}\cos (en\beta \tilde{A}_{3})\;,  
\label{bnboson}
\end{equation}
\noindent where $n=1,2,...$ while 
$b_{0}=-e\beta B(m+\omega _{c})/2\pi$ does not contribute to the 
thermodynamics. In analogy with the fermionic case we can write
the cluster coefficients here as  
\begin{equation}
{b_{n}^{B}}|_{LLL}=\frac{\rho _{L}}{n}
e^{-n\beta (m+\omega _{c})}\cos (n\theta)\; .
\end{equation}

Then, we find that the virial coefficients in this case are given 
by 
\begin{eqnarray}  
\label{aB2}
a_{2}^B|_{LLL} &=& \;-\; \frac{1}{2\rho _{L}}[1-\tan ^{2}\theta ]; \\
a_{3}^B|_{LLL} &=& \;+\; \frac{1}{3(\rho _{L})^{2}}[1+3\tan ^{4}\theta ]; \\
a_{4}^B|_{LLL} &=& \;-\; \frac{1}{4(\rho _{L})^{3}}[1-3\tan ^{4}{\theta }
+10\tan ^{6}{\ \theta }]\;; \\
a_{5}^B|_{LLL} &=& \;-\; \frac{1}{5(\rho _{L})^{4}}[1+20\tan ^{6}{\theta }
+35\tan ^{8}{\ \theta }]\;; \\
a_{6}^B|_{LLL} &=& \;+\; {\frac{1}{6(\rho _{L})^{5}}}
[1-10\tan ^{6}\theta -105\tan^{8}\theta +126\tan ^{10}\theta ]\;.
\end{eqnarray}

\noindent These coefficients are closed related with those obtained 
in the fermionic case, namely: 
\[
a_{n}^B|_{LLL}=(-1)^{n-1}a_{n}^{LLL} \; . 
\]

They may interpolate between the bosonic and fermionic coefficients 
as happens for particles without magnetic field background \cite{bbf123}. 
An analogous situation happens in the fermionic case discussed previously.
These coefficients differ in general from those of anyons 
\cite{leinaas77,arovas85,JC90,kim,VO94}. Note that the bosonic virial
coefficients are also singular for $\theta =\pi/2$, which seems to 
prevent spontaneous transmutation of bosons into fermions and 
vice-versa.

%%%%%%%%%%%%%%%%%%%%%%%%%%%%%%%%%%%%%%%%%%%%%%%%%%%%%%%%%%%%%%%%%
%%%%%%%%%%%%%%%%%%%%%%%%%%%%%%%%%%%%%%%%%%%%%%%%%%%%%%%%%%%%%%%%%

\section{Conclusions}

Virial coefficients of a gas of relativistic particles can be 
obtained from the fundamental quantum theory of electromagnetic 
interactions, \textsl{i. e. }, QED, which guarantees large gauge 
invariance in 2+1 dimensions at finite temperature and density. 
These depend on an arbitrary parameter $\theta $ that carries 
a dependence on the temperature and the time-component of the 
gauge field. We have shown that such parameter could play the 
role of an interpolation parameter between the bosonic and 
fermionic virial coefficients, although these virial coefficients
are singular for $\theta =\pi/2$.

It is important to stress that the approach used here in the 
calculation of the virial coefficients is based on finite 
temperature and density effects of the fundamental bosonic 
and fermionic particles, similar to that discussed in the 
absence of magnetic fields \cite{bbf123} and suggested in 
the literature \cite{barbosa}. 
In other words, the interpolation discussed here appears only 
at finite temperature and density, since at zero temperature 
large gauge transformations become trivial and the (constant) 
parameter associated with the time component of the gauge field 
can be removed by an ordinary (infinitesimal) gauge 
transformation.

Finally, in this work we have not discussed interactions of the 
fundamental fermions (or bosons) with other fields than the 
constant magnetic background. This implied that the constant 
parameter $\theta$ (or $\tilde{A}_3$) was not fixed here. 
We expect that the basic characteristics discussed here 
related to finite temperature and density interpolation between  
bosonic and fermionic virial coefficients would survive to the 
presence of other interactions. 

\bigskip

\noindent {\bf Acknowledgments.}
 We thank Y. S. Myung for calling our attention to refs. \cite{kim}. 
H.B.-F. was partially supported by CNPq and CAPES/PROCAD 
and M.H. by FAPESP (Brazilian agencies).

\end{document}